\documentstyle[12pt]{article}
\makeatletter
\newcommand{\singlespacing}{\let\CS=\@currsize\renewcommand{\baselinestretch}
{1.0}\tiny\CS}
\newcommand{\doublespacing}{\let\CS=\@currsize\renewcommand{\baselinestretch}
{1.5}\tiny\CS}

\begin{document}

\begin{center}
{\Large {\bf Constant rotation of two-qubit equally entangled pure
states by local quantum operations}}

\vspace{1cm} {\large{\bf Samir Kunkri}}$^a$, {\large{\bf Swarup
Poria}}$^b$, {\large{\bf Preeti Parashar}}$^c$ {\large and}\\
{\large{\bf Sibasish Ghosh}}$^d$

\vspace{1cm} $^a${\large{\it Mahadevananda Mahavidyalaya,
Monirampore, Barrackpore, North 24 Parganas, 700120, West Bengal,
India.}}\footnote{\verb"E-mail: skunkri@yahoo.com"}

\vspace{0.5cm} $^b${\large{\it Department of Mathematics, Midnapore
College, Midnapore, 721101, West Bengal,
India.}}\footnote{\verb"E-mail: swarup"$_-$\verb"p@yahoo.com"}

\vspace{0.5cm} $^c${\large{\it Physics and Applied Mathematics Unit,
Indian Statistical Institute, 203 B. T. Road, Kolkata 700108,
India.}}\footnote{\verb"E-mail: parashar@isical.ac.in"}

\vspace{0.5cm} $^d${\large{\it The Institute of Mathematical
Sciences, C. I. T. Campus, Taramani, Chennai 600113,
India.}}\footnote{\verb"E-mail: sibasish@imsc.res.in"}

\end{center}

\date{today}

\vspace{0.5cm}

\begin{center}
{\bf Abstract}
\end{center}

{\small We look for local unitary operators $W_1 \otimes W_2$ which
would rotate all equally entangled two-qubit pure states by the same
but arbitrary amount. It is shown that all two-qubit maximally
entangled states can be rotated through the same but arbitrary
amount by local unitary operators. But there is no local unitary
operator which can rotate all equally entangled non-maximally
entangled states by the same amount, unless it is unity. We have
found the optimal sets of equally entangled non-maximally entangled
states which can be rotated by the same but arbitrary amount via
local unitary operators $W_1 \otimes W_2$, where at most one these
two operators can be identity. In particular, when $W_1 = W_2 =
(i/\sqrt{2})({\sigma}_x + {\sigma}_y)$, we get the local quantum NOT
operation. Interestingly, when we apply the one-sided local
depolarizing map, we can rotate all equally entangled two-qubit pure
states through the same amount. We extend our result for the case of
three-qubit maximally entangled state.}

\section{Introduction}
Fundamental limitation of certain quantum operations of single
systems with regards to global operations, has already been studied
in the literature \cite{wootters, pati1, unot, pati2}. Cloning and
the NOT operation, applied to the qubit, are the two prime example
\cite{wootters, unot}. It is well known that an arbitrary state,
taken from a set of two known non-orthogonal states, can not be
copied exactly. But any state taken from a set of orthogonal states
can be copied exactly. Similarly, there exits no universal flipper
which can operate on a unknown qubit state $|\psi\rangle$, resulting
in the orthogonal state $|\psi^{\perp}\rangle$. The largest set of
states (of single-qubit system), which can be flipped exactly by a
single unitary operator, is the set of states lying on a great
circle of the Bloch sphere \cite{pati2}.

Entanglement lies at the heart of many aspects of quantum
information theory and it is therefore desirable to understand its
structure as well as possible \cite{chuang}. One attempt to improve
our understanding of entanglement is the study of our ability to
perform information theoretic tasks locally on entangled state, such
as local discrimination, local cloning etc. \cite{hardy00, ghosh01,
ghosh04, plenio}. It has already been established that these local
scenarios of quantum operations are different from the global
scenarios. For example, in the global scenario any state given from
the set of orthogonal states can be cloned exactly, whereas there
are examples of set of orthogonal entangled states which can not be
cloned exactly by local operation \cite{ghosh04, plenio}.

Recently Novotn$\acute{{\rm y}}$ et al. \cite{jex} studied the
optimal covariant\footnote{If a completely positive map $T : {\cal
B}({C\!\!\!\!I}^2 \otimes {C\!\!\!\!I}^2) \rightarrow {\cal
B}({C\!\!\!\!I}^2 \otimes {C\!\!\!\!I}^2)$ has to treat pure
two-qubit density matrices $\rho$, of given degree of entanglement,
in a {\it covariant} way, one must have $T((U \otimes
V){\rho}(U^{\dagger} \otimes V^{\dagger})) = (U \otimes
V)T(\rho)(U^{\dagger} \otimes V^{\dagger})$ for all $U, V \in
SU(2)$.} quantum NOT operations for equally entangled qubit pairs,
in the global scenario. In particular, they have shown that only in
the case of maximally entangled input states, such covariant quantum
NOT operations can be performed perfectly. In the case of maximally
entangled states of two qubits, they have also discussed about
universal non-covariant quantum NOT operations. Motivated by these
studies, and in the line of the tasks of local cloning, local
deleting, local broadcasting, locally distinguishing, etc., we would
like to investigate here the possibility of locally rotating,
through the same amount, equally entangled two-qubit states. We here
consider implementation of an exact quantum NOT operation, and its
generalization (namely, rotation) by using LOCC only. In particular,
we will show here that {\it all} two-qubit maximally entangled
states can be exactly rotated through the same (but arbitrarily
given) amount by the action of an one-sided local unitary operator.
We also show that for given an arbitrary value of two-qubit pure
entanglement ($E$, say), there exists a maximal subset of the set of
all two-qubit pure states having entanglement $E$, such that all the
elements of that maximal set can be exactly rotated through the same
(but arbitrarily given) amount by the action of an one-sided local
unitary operator\footnote{Throughout this paper, we take only those
unitary matrices each of whose determinant is unity. But the results
of this paper also follows for unitary operators having determinant
other than unity -- simply we need to multiply the respective
special unitary operators by a phase.}. It is shown that in the case
of equally entangled non-maximally entangled two-qubit pure states,
not all the states in the whole class of such states can be rotated
through a constant amount by local unitary operation. But in the
case of maximally entangled states, all the states in this class can
be rotated through a constant amount by local unitary operation. As
a consequence of this general studies we will discuss the local-NOT
operation of pure two-qubit states with a fixed degree of
entanglement.

In section 2, we describe the general scheme of `rotating', via same
amount, pure states of $2 \otimes 2$ having same amount of
entanglement, by local unitary operators $W_1 \otimes W_2$, where
$W_1, W_2 \in SU(2)$. In section 3, we consider the issue of
rotation, through the same amount' of equally entangled two-qubit
pure states by one-sided unital trace-preserving completely (CP)
positive maps. In section 4, we consider the issue of rotating, via
same amount, maximally entangled states of three qubits by local
unitary operators. We shall draw the conclusion in section 5.

\section{Constant rotation of equally entangled two-qubit pure states by local unitary operators}
To start with, we consider a two-qubit pure state
\begin{equation}
\label{psi0} |{\psi}_0\rangle = l_0|00\rangle + l_1|11\rangle,
\end{equation}
having Schmidt coefficients $l_0$, $l_1$ (with $0 \le l_1 \le l_0
\le 1$ and $l_0^2 + l_1^2 = 1$) and Schmidt basis $\{|0\rangle_i,
|1\rangle_i\}$ for the $i$-th particle ($i = 1, 2$). Here the given
amount of entanglement is $E = - l_0^2 {\rm log}_2 l_0^2 - l_1^2
{\rm log}_2 l_1^2$. Let ${\cal S}_{(l_0, l_1)} = \{(U \otimes
V)|{\psi}_0\rangle :~ U, V \in SU(2)\}$ be the collection of {\it
all} the two-qubit pure states, each having entanglement $E$.

\subsection{Constant rotation of equally entangled two-qubit pure
states by one-sided local unitary operators} Let us now look for an
one-sided local unitary operator $W_1 \otimes I_2 \in SU(2) \otimes
SU(2)$ (or $I_2 \otimes W_2 \in SU(2) \otimes SU(2)$) which can
rotate $|{\psi}_0\rangle$ as well as some or all the other elements
of ${\cal S}_{(l_0, l_1)}$ through one and the same given amount
$re^{i\theta}$ (with $0 \le r \le 1$ and $0 \le \theta \le 2\pi$).
Taking $W_1 = r_0^{W_1}I_2 + i\overrightarrow{r^{W_1}}.\vec{\sigma}$
(with $(r_0^{W_1}, \overrightarrow{r^{W_1}}) \in {I\!\!\!\!R}^2$ and
$(r_0^{W_1})^2 + |\overrightarrow{r^{W_1}}|^2 = 1$), we get from the
condition ${\langle}{\psi}_0|W_1 \otimes I_2|{\psi}_0{\rangle} =
re^{i\theta}$ that $r_0^{W_1} = r {\rm cos} \theta$ and
$r_z^{W_1}(l_0^2 - l_1^2) = r {\rm sin} \theta$. Thus, we are
looking for the maximal subset ${\cal S}^{(l_0, l_1)}_{W_1 \otimes
I_2}$ of states $(U \otimes V)|{\psi}_0\rangle$ from ${\cal
S}_{(l_0, l_1)}$ for which ${\langle}{\psi}_0|U^{\dagger}W_1U
\otimes I_2|{\psi}_0{\rangle} = re^{i\theta} \equiv r_0^{W_1} +
ir_z^{W_1}(l_0^2 - l_1^2)$. Now $U^{\dagger}W_1U = r_0^{W_1}I_2 +
i(R_{U^{\dagger}}\overrightarrow{r^{W_1}}).\vec{\sigma}$, where, for
any $U \in SU(2)$, $R_{U}$ is the $3 \times 3$ real orthogonal
rotation matrix corresponding to $U$. In other words, for any
element $\vec{a} \in {I\!\!\!\!R}^3$, we have
\begin{equation}
\label{RU} R_U\vec{a} \equiv
2(\overrightarrow{r^U}.\vec{a})\overrightarrow{r^U} + (1 -
2|\overrightarrow{r^U}|^2)\vec{a} - 2r_0^U(\overrightarrow{r^U}
\times \vec{a}).
\end{equation}
So we have ${\langle}{\psi}_0|U^{\dagger}W_1U \otimes
I_2|{\psi}_0{\rangle} = r_0^{W_1} +
i(R_{U^{\dagger}}\overrightarrow{r^{W_1}})_z(l_0^2 - l_1^2)$. Thus
we are looking for the {\it maximal} set of $(U \otimes V)$'s from
$SU(2) \otimes SU(2)$ so that for each such $U \otimes V$, we have
\begin{equation}
\label{W1} (l_0^2 -
l_1^2)[(R_{U^{\dagger}}\overrightarrow{r^{W_1}})_z - r_z^{W_1}] = 0.
\end{equation}
If $|{\psi}_0\rangle$ is a maximally entangled state, condition
(\ref{W1}) is thus satisfied for {\it all} $U, V \in SU(2)$, {\it
i.e.}, for all two-qubit maximally entangled states. If
$|{\psi}_0\rangle$ is not a maximally entangled state, then the
desired maximal set will consist of only those states $(U \otimes
V)|{\psi}_0\rangle$ for which $V$ can be an {\it arbitrary} element
of $SU(2)$ but $U$ will be such that Bloch vectors
$R_{U^{\dagger}}(\overrightarrow{r^{W_1}}/|\overrightarrow{r^{W_1}}|)$
span a small circle of the Bloch sphere whose plane is perpendicular
to $\hat{z}$ and is at a distance
$r_z^{W_1}/|\overrightarrow{r^{W_1}}|$ from the centre of the Bloch
sphere. So $R_{U^{\dagger}}$ corresponds to the unitary operator
${\rm exp} [- i(\theta/2){\sigma}_z] =~ {\rm cos} (\theta/2) I_2 - i
{\rm sin} (\theta/2) {\sigma}_z$, {\it i.e.}, rotation about the
$z$-axis through some angle $\theta$.

Rotation by one-sided local unitary operators of the form $I_2
\otimes W_2$ will provide the similar result.

Note that by a single unitary operator $U \in SU(2)$, one can
rotate, through the same angle, only those single-qubit pure states
whose Bloch vectors lie on a particular (depending upon $U$) circle
of the Bloch sphere.

\subsection{Constant rotation of two-qubit pure states by one-sided
local unitary operators} Another way to rotate two-qubit pure states
by one-sided local unitaries of the form $(W_1 \otimes I_2) \in
SU(2) \otimes SU(2)$, through the same amount, is to first consider
the set ${\cal S}(W_1)$ of all single-qubit pure states
$|{\psi}(\theta, \phi)\rangle \equiv~ {\rm cos} (\theta/2) |0\rangle
+ e^{i\phi} {\rm sin} (\theta/2) |1\rangle \equiv U|0\rangle \equiv~
({\rm cos} (\theta/2) I_2 + i {\rm sin} \phi {\sin} (\theta/2)
{\sigma}_x - i {\rm cos} \phi {\rm sin} (\theta/2) {\sigma}_y)
|0\rangle$, each of which can be rotated through the same amount
${\langle}0|W_1|0{\rangle} = r_0^{W_1} + r_z^{W_1}$ by the unitary
matrix $W_1$. Note that
$${\cal S}(W_1) = \left\{|{\psi}(\theta,
\phi)\rangle :~ \hat{v}(\theta, \phi).{\overrightarrow{r^{W_1}}} =
r_z^{W_1}\right\},$$ where $\hat{v}(\theta, \phi) \equiv ({\rm sin}
\theta~ {\rm cos} \phi,~ {\rm sin} \theta~ {\rm sin} \phi,~ {\rm
cos} \theta)$ is the Bloch vector of the state $|\psi(\theta,
\phi)\rangle$. Thus ${\cal S}(W_1)$ is a circular section
(generally, a small circle) of the Bloch sphere, perpendicular to
the vector $\overrightarrow{r^{W_1}}$, while the projection of each
Bloch vector of that circle along $\overrightarrow{r^{W_1}}$ is same
as $r_z^{W_1}$. It is then easy to show that all the members of the
following set of two-qubit pure states
$${\cal S}_{W_1} \equiv \left\{|\chi\rangle =
\sqrt{\lambda}\left|{\psi}\left({\theta}_1,
{\phi}_1\right)\right\rangle \otimes |e\rangle + \sqrt{1 -
\lambda}\left|{\psi}\left({\theta}_2, {\phi}_2\right)\right\rangle
\otimes |e^{\bot}\rangle~ : \left|{\psi}\left({\theta}_1,
{\phi}_1\right)\right\rangle, \left|{\psi}\left({\theta}_2,
{\phi}_2\right)\right\rangle \in\right.$$
$$\left.{\cal S}(W_1),~ \lambda
\in [0, 1],~ \{|e\rangle, |e^{\bot}\rangle\}~ {\rm is}~ {\rm an}~
{\rm ONB}~ {\rm of}~ {C\!\!\!\!I}^2\right\}$$ can be rotated through
the same amount $r_0^{W_1} + r_z^{W_1}$ by the one-sided local
unitary $(W_1 \otimes I_2)$. Note that the single-qubit reduced
density matrix ${\rho}^{\chi}_1 \equiv~ {\rm Tr}_2
[|\chi{\rangle}{\langle}\chi|] = (1/2)[I_2 +
\{{\lambda}\hat{v}({\theta}_1, {\phi}_1) + (1 -
\lambda)\hat{v}({\theta}_2, {\phi}_2)\}.\vec{\sigma}]$ of
$|\chi\rangle$ has eigen values $(1 \pm |\vec{R}|)/2$, where
$|\vec{R}|^2 \equiv |{\lambda}\hat{v}({\theta}_1, {\phi}_1) + (1 -
\lambda)\hat{v}({\theta}_2, {\phi}_2)|^2 = {\lambda}^2 + (1 -
\lambda)^2 + 2{\lambda}(1 - \lambda)(\hat{v}({\theta}_1,
{\phi}_1).\hat{v}({\theta}_2, {\phi}_2)) = 1 - 2{\lambda}(1 -
{\lambda})(1 - \hat{v}({\theta}_1, {\phi}_1).\hat{v}({\theta}_2,
{\phi}_2))$. {\it So, not all the states $|\chi\rangle \in {\cal
S}_{W_1}$ are equally entangled.} In the special case when
$\hat{v}({\theta}_1, {\phi}_1).\hat{v}({\theta}_2, {\phi}_2) = - 1$
(which happens only when ${\cal S}(W_1)$ is a great circle, {\it
i.e.}, when $W_1$ is a NOT operation), for arbitrary but fixed
$\lambda \in [0, 1]$, each member of the set of all the {\it equally
entangled} two-qubit pure states $\sqrt{\lambda}|{\psi}({\theta},
{\phi})\rangle \otimes |e\rangle + \sqrt{1 - \lambda}|{\psi}(\pi -
{\theta}, \pi + {\phi})\rangle \otimes |e^{\bot}\rangle$ can be
rotated to its orthogonal state by the one-sided local unitary $(W_1
\otimes I_2)$, where $\{|e\rangle, |e^{\bot}\rangle\}$ is any ONB of
${C\!\!\!\!I}^2$. For any given Schmidt coefficients $l_0 = \sqrt{1
- {\lambda}(1 - {\lambda})(1 - \hat{v}({\theta}_1,
{\phi}_1).\hat{v}({\theta}_2, {\phi}_2))}$ and $l_1 =
\sqrt{{\lambda}(1 - {\lambda})(1 - \hat{v}({\theta}_1,
{\phi}_1).\hat{v}({\theta}_2, {\phi}_2))}$, all the members of the
subset (${{\cal S}^{\prime}}^{(l_0, l_1)}_{W_1}$ (say) of ${\cal
S}_{W_1}$) of all the two-qubit pure states $|\chi\rangle$, each
having one and the same pair of Schmidt coefficients $(l_0, l_1)$,
can be rotated through the same amount ${\langle}0|W_1|0{\rangle}$
by the local unitary operator $W_1 \otimes I_2$, where as, all the
members of the set ${\cal S}^D_{W_1 \otimes I_2} =
\{|UDV^T{\rangle}\!{\rangle} :~ V \in SU(2)~ {\rm and}~ T_{1z} =
0\}$, with $D =~ {\rm diag} (l_0, l_1)$ and
$$\vec{T}_1 =
\left|\overrightarrow{r^U}\right|^2\overrightarrow{r^{W_1}} -
\left(\overrightarrow{r^U}.\overrightarrow{r^{W_1}}\right)\overrightarrow{r^U}
- r^U_0\left(\overrightarrow{r^U} \times
\overrightarrow{r^{W_1}}\right),$$ can be rotated\footnote{For
definition of $|UDV^T{\rangle}\!{\rangle}$, please look at the last
but one paragraph of the conclusion section.} by $W_1 \otimes I_2$
through the same amount ${\langle}{\psi}_0|(W_1 \otimes
I_2)|{\psi}_0{\rangle}$ (where $|{\psi}_0\rangle = l_0|00\rangle +
l_1|11\rangle$). By mere parameter counting, one can see that
${{\cal S}^{\prime}}^{(l_0, l_1)}_{W_1}$ can be specified by {\it at
most four independent real parameters}, while ${\cal S}^D_{W_1
\otimes I_2}$ can be specified by {\it at most five independent real
parameters}. Hence, from this perspective, the size of ${{\cal
S}^{\prime}}^{(l_0, l_1)}_{W_1}$ seems to be smaller than that of
${\cal S}^D_{W_1 \otimes I_2}$.

\subsection{Constant rotation of equally entangled two-qubit pure
states by two-sided local unitary operators} Now we come to the
issue of enlarging the maximal set ${\cal S}^{(l_0, l_1)}_{W_1
\otimes I_2}$ (given by equation (\ref{W1})) by considering
rotations using both-sided local unitary operators $W_1 \otimes W_2
\in SU(2) \otimes SU(2)$. Both-sided local unitary operators, in
general, give rise to maximal sets whose sizes are different from
that for one-sided local unitary operations.

In fact, here we look for the maximal subset ${\cal S}^{(l_0,
l_1)}_{W_1 \otimes W_2}$ of the elements $(U \otimes
V)|{\psi}_0\rangle$ of ${\cal S}_{(l_0, l_1)}$ for which
${\langle}{\psi}_0|U^{\dagger}W_1U \otimes
V^{\dagger}W_2V|{\psi}_0{\rangle} = {\langle}{\psi}_0|W_1 \otimes
I_2|{\psi}_0{\rangle}$, {\it i.e.},
$$r_0^{W_1}(R_{V^{\dagger}}\overrightarrow{r^{W_2}})_z +
r_0^{W_2}(R_{U^{\dagger}}\overrightarrow{r^{W_1}})_z +
2l_0l_1\{(R_{U^{\dagger}}\overrightarrow{r^{W_1}})_x(R_{V^{\dagger}}\overrightarrow{r^{W_2}})_x
-
(R_{U^{\dagger}}\overrightarrow{r^{W_1}})_y(R_{V^{\dagger}}\overrightarrow{r^{W_2}})_y\}
+
(R_{U^{\dagger}}\overrightarrow{r^{W_1}})_z(R_{V^{\dagger}}\overrightarrow{r^{W_2}})_z$$
\begin{equation}
\label{W1W2} = r_0^{W_1}r^{W_2}_z + r_0^{W_2}r^{W_1}_z +
2l_0l_1\{r^{W_1}_xr^{W_2}_x - r^{W_1}_yr^{W_2}_y\} +
r^{W_1}_zr^{W_2}_z.
\end{equation}

To get an idea about the maximal set, let us consider the example
where $W_1 = W_2 = (i/\sqrt{2})({\sigma}_x + {\sigma}_y)$. So the
above-mentioned condition (\ref{W1W2}) now becomes
$2l_0l_1\{(R_{U^{\dagger}}\overrightarrow{r^{W_1}})_x(R_{V^{\dagger}}\overrightarrow{r^{W_2}})_x
-
(R_{U^{\dagger}}\overrightarrow{r^{W_1}})_y(R_{V^{\dagger}}\overrightarrow{r^{W_2}})_y\}
+
(R_{U^{\dagger}}\overrightarrow{r^{W_1}})_z(R_{V^{\dagger}}\overrightarrow{r^{W_2}})_z
= 0$, which, in turn, implies that the two vectors $\vec{T}_1(U;
l_0, l_1) \equiv
2l_0l_1(R_{U^{\dagger}}\overrightarrow{r^{W_1}})_x\hat{x} -
2l_0l_1(R_{U^{\dagger}}\overrightarrow{r^{W_1}})_y\hat{y} +
(R_{U^{\dagger}}\overrightarrow{r^{W_1}})_z\hat{z} \equiv
2l_0l_1(R_{U^{\dagger}}\overrightarrow{r^{W_1^T}})_x\hat{x} +
2l_0l_1(R_{U^{\dagger}}\overrightarrow{r^{W_1^T}})_y\hat{y} +
(R_{U^{\dagger}}\overrightarrow{r^{W_1^T}})_z\hat{z}$ and
$\vec{T}_2(V) \equiv R_{V^{\dagger}}\overrightarrow{r^{W_2}}$ are
orthogonal. In other words, for arbitrarily given $V \in SU(2)$, $U
\in SU(2)$ should be such that the Bloch vector $\vec{T}_1(U; l_0,
l_1)/|\vec{T}_1(U; l_0, l_1)|$ lies on the great circle (${\cal
G}(\vec{T}_2(V))$, say), orthogonal to the Bloch vector
$\vec{T}_2(V)/|\vec{T}_2(V)|$. So, for every $V \in SU(2)$,
$R_{U^{\dagger}}\overrightarrow{r^{W_1^T}}$ will an arbitrary Bloch
vector lying on the great circle perpendicular to the Bloch vector
$R_{V^{\dagger}}\overrightarrow{r^{W_2}}$. In other words, for every
$V \in SU(2)$, $R_{U^{\dagger}}$ implies rotation (through some
angle) about the vector $R_{V^{\dagger}}\overrightarrow{r^{W_2}}$.
Thus we see that the size of the required maximal set ${\cal
S}^{(l_0, l_1)}_{W_1 \otimes W_2}$ is {\it same} as in the case of
one-sided local unitary operations.

\section{Constant rotation of equally entangled two-qubit pure
states by one-sided unital trace-preserving CP maps} Considering the
action of the local quantum operations at the density matrix level,
we may look for the maximal set ${\cal T}^{(l_0, l_1)}_{W_1 \otimes
W_2} \equiv \{(U \otimes V)|{\psi}_0\rangle \in {\cal S}_{(l_0,
l_1)} :~ {\langle}{\psi}_0|(U^{\dagger} \otimes V^{\dagger})\{(W_1
\otimes W_2)((U \otimes
V)|{\psi}_0{\rangle}{\langle}{\psi}_0|(U^{\dagger} \otimes
V^{\dagger}))(W_1^{\dagger} \otimes W_2^{\dagger})\}(U \otimes
V)|{\psi}_0{\rangle} = {\langle}{\psi}_0|\{(W_1 \otimes
W_2)|{\psi}_0{\rangle}{\langle}{\psi}_0|(W_1^{\dagger} \otimes
W_2^{\dagger})\}|{\psi}_0{\rangle}\} = \{(U \otimes
V)|{\psi}_0\rangle \in {\cal S}_{(l_0, l_1)} :~
|{\langle}{\psi}_0|U^{\dagger}W_1U \otimes
V^{\dagger}W_2V|{\psi}_0{\rangle}|^2 = |{\langle}{\psi}_0|W_1
\otimes W_2|{\psi}_0{\rangle}|^2\}$, instead of the maximal set
${\cal S}^{(l_0, l_1)}_{W_1 \otimes W_2}$. This immediately shows
that ${\cal S}^{(l_0, l_1)}_{W_1 \otimes W_2} \subset {\cal
T}^{(l_0, l_1)}_{W_1 \otimes W_2}$. Thus, for example, ${\cal
T}^{(l_0, l_1)}_{W_1 \otimes I_2}$ will consist of all the states
$(U \otimes V)|{\psi}_0\rangle \in {\cal S}_{(l_0, l_1)}$ where $V$
is an arbitrary element of $SU(2)$ while $U \in S(2)$ is such that
$R_{U^{\dagger}}$ can be either a rotation about the positive $z$
axis or a rotation about the negative $z$ axis.

We now raise the issue of further enlargement of the maximal set
${\cal T}^{(l_0, l_1)}_{W_1 \otimes W_2}$ by using LOCC, not just
local unitary operators. Here, for simplicity, we consider only the
question of rotating, through the same amount, all or some of the
elements of ${\cal S}_{(l_0, l_1)}$ by using an one-sided local
unital CP map $T_1 \otimes I_2$, where $T_1$ is a unital,
trace-preserving CP map on ${\cal B}({C\!\!\!\!I}^2)$. Any unital
trace-preserving CP map $T_1 : {\cal B}({C\!\!\!\!I}^2) \rightarrow
{\cal B}({C\!\!\!\!I}^2)$ can be expressed as $T_1(\rho) = \sum_{j =
1}^{4} {\lambda}_jW_j^{\prime}{\rho}{W_j^{\prime}}^{\dagger}$ where
$\rho \in {\cal B}({C\!\!\!\!I}^2)$, $W_j^{\prime} \in SU(2)$, $0
\le {\lambda}_j \le 1$ (for $j = 1, 2, 3, 4$), and $\sum_{j = 1}^{4}
{\lambda}_j = 1$ \cite{ruskai01}. Thus we are looking for the
maximal subset ${\cal T}^{(l_0, l_1)}_{T_1 \otimes I_2}$ of states
$(U \otimes V)|{\psi}_0\rangle$ (including the state
$|{\psi}_0\rangle$) from ${\cal S}_{(l_0, l_1)}$ for which
$$\sum_{i = 1}^{4} {\lambda}_i{\langle}{\psi}_0|(U^{\dagger}W_i^{\prime}U \otimes
I_2)|{\psi}_0{\rangle}{\langle}{\psi}_0|(U^{\dagger}W_i^{\prime}U
\otimes I_2)^{\dagger}|{\psi}_0{\rangle}$$
$$= \sum_{i = 1}^{4}
{\lambda}_i~ {\langle}{\psi}_0|(W_i^{\prime} \otimes
I_2)|{\psi}_0{\rangle}{\langle}{\psi}_0|(W_i^{\prime} \otimes
I_2)^{\dagger}|{\psi}_0{\rangle}.$$ We are thus looking for all
those $U, V \in SU(2)$ for which
$$l_0^2\sum_{j = 1}^{4}
{\lambda}_j|{\langle}{\psi}_0|\{((U^{\dagger}W_j^{\prime}U)|0\rangle)
\otimes |0\rangle\}|^2 + l_1^2\sum_{j = 1}^{4}
{\lambda}_j|{\langle}{\psi}_0|\{((U^{\dagger}W_j^{\prime}U)|1\rangle)
\otimes |1\rangle\}|^2 +$$
$$2l_0l_1\sum_{j = 1}^{4} {\lambda}_j~ {\rm Re}
([{\langle}{\psi}_0|\{((U^{\dagger}W_j^{\prime}U)|0\rangle) \otimes
|0\rangle\}] \times
[{\langle}{\psi}_0|\{((U^{\dagger}W_j^{\prime}U)|1\rangle) \otimes
|1\rangle\}]^*) = l_0^2\sum_{j = 1}^{4}
{\lambda}_j|{\langle}{\psi}_0|\{(W_j^{\prime}|0\rangle) \otimes
|0\rangle\}|^2 +$$
$$l_1^2\sum_{j = 1}^{4}
{\lambda}_j|{\langle}{\psi}_0|\{(W_j^{\prime}|1\rangle) \otimes
|1\rangle\}|^2 + 2l_0l_1\sum_{j = 1}^{4} {\lambda}_j~ {\rm Re}
([{\langle}{\psi}_0|\{(W_j^{\prime}|0\rangle) \otimes |0\rangle\}]
\times [{\langle}{\psi}_0|\{(W_j^{\prime}|1\rangle) \otimes
|1\rangle\}]^*),$$ which is nothing but the following condition:
\begin{equation}
\label{T1cond} (l_0^2 - l_1^2)^2\sum_{j = 1}^{4}
{\lambda}_j\left[\left\{\left(R_{U^{\dagger}}\overrightarrow{r^{W_j^{\prime}}}\right)_z\right\}^2
- \left(r^{W_j^{\prime}}_z\right)^2\right] = 0.
\end{equation}
For a maximally entangled state $|{\psi}_0\rangle$, this condition
is {\it automatically satisfied for all} $U, V \in SU(2)$. When
$|{\psi}_0\rangle$ is a non-maximally entangled state, the
above-mentioned condition becomes
\begin{equation}
\label{T1condnonmax} \sum_{j = 1}^{4}
{\lambda}_j\left[\left\{\left(R_{U^{\dagger}}\overrightarrow{r^{W_j^{\prime}}}\right)_z\right\}^2
- \left(r^{W_j^{\prime}}_z\right)^2\right] = 0.
\end{equation}
If $R_{U^{\dagger}}$ represents a rotation about the $z$-axis of the
Bloch sphere, then
$(R_{U^{\dagger}}\overrightarrow{r^{W_j^{\prime}}})_z =
r_z^{W_j^{\prime}}$ for $j = 1, 2, 3, 4$, and so, condition
(\ref{T1condnonmax}) will be satisfied in this case. We have seen
this solution earlier in the case of constant rotation by one-sided
local unitary transformation $W_1 \otimes I_2$. But equation
(\ref{T1condnonmax}) will have more solution(s). This implies that
the size of ${\cal T}^{(l_0, l_1)}_{W_1 \otimes I_2}$ will be {\it
smaller}, in general, than the size of set ${\cal T}^{(l_0,
l_1)}_{T_1 \otimes I_2}$.

In the case of the bit-flip channel $T_1(\rho) \equiv p{\rho} + (1 -
p){\sigma}_x{\rho}{\sigma}_x$ (with $0 \le p < 1$), we have
${\lambda}_1 = p$, ${\lambda}_2 = (1 - p)$, ${\lambda}_3 =
{\lambda}_4 = 0$ and $W_1^{\prime} = I_2$, $W_2^{\prime} =
i{\sigma}_x$. Then the above-mentioned condition becomes $(1 -
p)\{(R_{U^{\dagger}}\hat{x})_z\}^2 = 0$, {\it i.e.},
$(R_{U^{\dagger}}\hat{x})_z = 0$. So, as in the case of equal
rotations by one-sided local unitary operators, here
$R_{U^{\dagger}}$ will correspond to rotations about the $z$-axis of
the Bloch sphere.

On the other hand, for the depolarizing channel $T_1(\rho) \equiv
p{\rho} + ((1 - p)/3)[{\sigma}_x{\rho}{\sigma}_x +
{\sigma}_y{\rho}{\sigma}_y + {\sigma}_z{\rho}{\sigma}_z]$ (with $0
\le p < 1$ and $1 - p$ is called the `depolarization coefficient'),
we have ${\lambda}_1 = p$, ${\lambda}_2 = {\lambda}_3 = {\lambda}_4
= (1 - p)/3$ and $W_1^{\prime} = I_2$, $W_2^{\prime} = i{\sigma}_x$,
$W_3^{\prime} = i{\sigma}_y$, $W_4^{\prime} = i{\sigma}_z$. Then the
above-mentioned condition becomes $1 =
\{(R_{U^{\dagger}}\hat{x})_z\}^2 + \{(R_{U^{\dagger}}\hat{y})_z\}^2
+ \{(R_{U^{\dagger}}\hat{z})_z\}^2$. This condition is {\it always
satisfied whatever be the unitary operator} $U \in SU(2)$ (use
equation (\ref{RU})). Thus we see that if we want to rotate, through
one and the same amount $r \in [0, 1)$ (with $r \ge (l_0^2 -
l_1^2)^2/3$), the state $|{\psi}_0\rangle$ as well as all the other
members $(U \otimes V)|{\psi}_0\rangle$ of ${\cal S}_{(l_0, l_1)}$
by some one-sided local unital trace-preserving CP map $(T_1 \otimes
I) : {\cal B}({C\!\!\!\!I}^2 \otimes {C\!\!\!\!I}^2) \rightarrow
{\cal B}({C\!\!\!\!I}^2 \otimes {C\!\!\!\!I}^2)$ (in the sense that
${\langle}{\psi}_0|(U^{\dagger} \otimes V^{\dagger})[(T_1 \otimes
I)((U \otimes V)|{\psi}_0{\rangle}{\langle}{\psi}_0|(U^{\dagger}
\otimes V^{\dagger}))](U \otimes V)|{\psi}_0{\rangle} =
{\langle}{\psi}_0|[(T_1 \otimes
I)(|{\psi}_0{\rangle}{\langle}{\psi}_0|)]|{\psi}_0{\rangle} \equiv
r$), we can do so by taking $T_1$ as the depolarizing channel with
$p = \{3r - (l_0^2 - l_1^2)^2\}/\{3 - (l_0^2 - l_1^2)^2\}$. So, when
$|{\psi}_0\rangle$ is not a maximally entangled state, the one-sided
local depolarizing channel $(T_1 \otimes I)$ {\it can not} act as a
NOT operation on the elements of set of all the equally entangled
states $(U \otimes V)|{\psi}_0\rangle$. Once again, {\it every}
maximally entangled state $|\psi\rangle$ of two qubits can be
transformed into a two-qubit state $\rho$, having support in the
orthogonal subspace of $|\psi\rangle$, by the one-sided local
depolarizing map $(T_1 \otimes I)$.

Note that by the action of the depolarizing map $T_1$ on all the
single-qubit pure states $U|0\rangle$ (where $U \in SU(2)$), every
such state can be state can be `rotated', through the same amount
${\langle}0|T_1(|0{\rangle}{\langle}0|)|0{\rangle} = (1 + 2p)/3$.
But this map can not act as a quantum NOT operation as here $1/3 \le
(1 + 2p)/3 < 1$.

\section{Local rotation of three-qubit maximally entangled states}
Let us consider the three-qubit GHZ state
\begin{equation}
\label{rotghz1} |GHZ\rangle = \frac{1}{\sqrt{2}}(|000\rangle +
|111\rangle).
\end{equation}
This is a maximally entangled state of three qubits. It is to be
noted that an $n$-qubit pure state is said to be a maximally
entangled state if and only if it has maximal available entanglement
in every bipartite cut. According to this definition, every
maximally entangled state $|{\psi}_{max}\rangle$ of three qubits is
connected to $|GHZ\rangle$ via local unitary of the form
$U_A(\alpha_1, \gamma_1, \delta_1) \otimes U_B(\alpha_2, \gamma_2,
\delta_2) \otimes U_C(\alpha_3, \gamma_3, \delta_3)$ and vice-versa,
where (with respect to the basis $\{|0\rangle, |1\rangle\}$)
\begin{equation}
\label{rotghz2} U_A(\alpha_1, \gamma_1, \delta_1) = \left(
\begin{array}{cc}
                                                   e^{i\delta_1}~
                                                   {\rm cos}
                                                   \alpha_1
                                                   & e^{i\gamma_1}~ {\rm
                                                   sin} \alpha_1\\
                                                   -e^{-i\gamma_1}~ {\rm
                                                   sin} \alpha_1 &
                                                   e^{-i\delta_1} ~
                                                   {\rm cos}
                                                   \alpha_1
                                                   \end{array}
                                                   \right),
\end{equation}
with $\alpha_1, \gamma_1, \delta_1 \in [0, 2\pi]$. Similar form for
the others unitary operators $U_B(\alpha_2, \gamma_2, \delta_2)$ and
$U_C(\alpha_3, \gamma_3, \delta_3)$. One can also show that
\begin{equation}
\label{rotghz3}
\begin{array}{lcl}
|{\psi}_{max}\rangle &=& \left(U(\alpha_1, \gamma_1, \delta_1)
\otimes U(\alpha_2, \gamma_2, \delta_2) \otimes U(\alpha_3,
\gamma_3, \delta_3)\right)|GHZ\rangle\\ &=& \left(I \otimes
V_{23}\left(\alpha_1, \gamma_1, \delta_1; \alpha_2, \gamma_2,
\delta_2; \alpha_3, \gamma_3, \delta_3 \right)\right)|GHZ\rangle,
\end{array}
\end{equation}
where $V_{23}(\alpha_1, \gamma_1, \delta_1; \alpha_2, \gamma_2,
\delta_2; \alpha_3, \gamma_3, \delta_3 )$ is a two-qubit unitary
matrix, acting on $|00\rangle$ and $|11\rangle$ in the following way
\begin{equation}
\label{rotghz4} \begin{array}{lcl} V_{23}|00\rangle &=&
e^{i\delta_1}~ {\rm cos} \alpha_1 \left|{\eta}\left(\alpha_2,
\gamma_2, \delta_2\right) {\eta}\left(\alpha_3, \gamma_3,
\delta_3\right)\right\rangle + e^{i\gamma_1}~ {\rm sin} \alpha_1
\left|{\overline{{\eta}\left(\alpha_2, \gamma_2, \delta_2\right)}}
{\overline{{\eta}\left(\alpha_3, \gamma_3,
\delta_3\right)}}\right\rangle,\\
V_{23}|11\rangle &=& -e^{-i\gamma_1}~ {\rm sin} \alpha_1
\left|{\eta}\left(\alpha_2, \gamma_2, \delta_2\right)
{\eta}\left(\alpha_3, \gamma_3, \delta_3\right)\right\rangle +
e^{-i\delta_1}~ {\rm cos} \alpha_1
\left|{\overline{{\eta}\left(\alpha_2, \gamma_2, \delta_2\right)}}
{\overline{{\eta}\left(\alpha_3, \gamma_3,
\delta_3\right)}}\right\rangle,
\end{array}
\end{equation}
with
\begin{equation}
\label{rotghz5}
\begin{array}{lcl}
\left|{\eta}\left(\alpha_2, \gamma_2, \delta_2\right)\right\rangle
&=& e^{i\delta_2}~ {\rm cos} \alpha_2|0\rangle - e^{-i\gamma_2}~
{\rm sin}
\alpha_2|1\rangle,\\
\left|{\overline{{\eta}\left(\alpha_2, \gamma_2,
\delta_2\right)}}\right\rangle &=& e^{i\gamma_2}~ {\rm sin}
\alpha_2|0\rangle + e^{-i\delta_2}~ {\rm cos} \alpha_2|1\rangle.
\end{array}
\end{equation}
Similar expression holds for the other states $|\eta (\alpha_3,
\gamma_3, \delta_3)\rangle$.

Given $\theta, \phi \in [0, 2\pi]$, let us now try to find out a
three-qubit quantum operation ${\cal A}$ such that
\begin{equation}
\label{rotghz6} \langle{\psi}_{max}|{\cal A}|{\psi}_{max}\rangle =
e^{i\phi}~ {\rm cos} \theta
\end{equation}
for {\it largest} number of three-qubit maximally entangled states
$|{\psi}_{max}\rangle$. One can {\it always} do so by choosing
$\alpha^{\prime}, \gamma^{\prime}, \delta^{\prime}$ in $[0, 2\pi]$
properly such that ${\cal A} = e^{i\phi}U(\alpha^{\prime},
\gamma^{\prime}, \delta^{\prime}) \otimes I \otimes I$, where
$U(\alpha^{\prime}, \gamma^{\prime}, \delta^{\prime})$ is given in
equation (\ref{rotghz2}) and $I$ is the single-qubit identity
operator. In fact, we have
\begin{equation}
\label{rotghz7} \langle{GHZ}|(e^{i\phi}U(\alpha^{\prime},
\gamma^{\prime}, \delta^{\prime}) \otimes I \otimes I)|{GHZ}\rangle
= e^{i\phi}~ {\rm cos} \alpha^{\prime}~ {\rm cos} \delta^{\prime}
\equiv e^{i\phi}~ {\rm cos} \theta.
\end{equation}
And (using equation (\ref{rotghz3}))
$$\left\langle{\psi}_{max}\right|(e^{i\phi}U(\alpha^{\prime}, \gamma^{\prime}, \delta^{\prime}) \otimes I
\otimes I)\left|{\psi}_{max}\right\rangle =$$
$$\langle{GHZ}|\left(e^{i\phi}U(\alpha^{\prime}, \gamma^{\prime}, \delta^{\prime}) \otimes
\left(V_{23}(\alpha_1, \gamma_1, \delta_1; \alpha_2, \gamma_2,
\delta_2; \alpha_3, \gamma_3,
\delta_3)\right)^{\dagger}V_{23}(\alpha_1, \gamma_1, \delta_1;
\alpha_2, \gamma_2, \delta_2; \alpha_3, \gamma_3,
\delta_3)\right)|{GHZ}\rangle$$
$$= \langle{GHZ}|(e^{i\phi}U(\alpha^{\prime}, \gamma^{\prime}, \delta^{\prime})
\otimes I \otimes I)|{GHZ}\rangle$$
\begin{equation}
\label{rotghz8} = e^{i\phi}~ {\rm cos} \alpha^{\prime}~ {\rm cos}
\delta^{\prime}.
\end{equation}
Thus we see that given $\theta, \phi \in [0, 2\pi]$, one can always
find out a single-qubit unitary matrix $U(\alpha^{\prime},
\gamma^{\prime}, \delta^{\prime})$ (given by equation
(\ref{rotghz2})) such that
$\langle{\psi}_{max}|(e^{i\phi}U(\alpha^{\prime}, \gamma^{\prime},
\delta^{\prime}) \otimes I \otimes I)|{\psi}_{max}\rangle =
e^{i\phi}~ {\rm cos} \theta$ for {\it all} three-qubit maximally
entangled states $|{\psi}_{max}\rangle$.

\section{Conclusion}
It is shown that all two-qubit pure maximally entangled states can
be rotated by a constant amount via local unitary operators of the
forms $W_1 \otimes I_2$ or $I_2 \otimes W_2$. But there exists no
local unitary operator which can rotate all non-maximally entangled
two-qubit pure states for a fixed degree of entanglement. We have
found the optimal sets of non-maximally entangled states for a fixed
degree of entanglement, each of which can be rotated through a
constant amount by local unitary operations of three different forms
$W_1 \otimes W_2$, $W_1 \otimes I_2$ and $I_2 \otimes W_2$. In
particular, when this amount is zero, we get the local quantum NOT
operation. Although the sizes of the optimal sets are, in general,
different for one-sided and two-sided local unitary operations, the
size of the optimal set for two-sided local quantum NOT operation is
same as that for any one-sided local unitary operation. Surprisingly
we found that the depolarizing map on one of the two qubits, all
equally entangled two-qubit pure states can be rotated through a
constant amount, whose value depends on the value of the
depolarization coefficient. Our result for two-qubits is extended
for the case of three-qubit maximally entangled state. Note that
Novotn$\acute{{\rm y}}$ et al. \cite{jex} have considered, most of
times, covariant trace-preserving two-qubit CP maps which would
serve as quantum NOT operations for the set of some or all two-qubit
equally entangled pure states. This covariance allows only the set
of all maximally entangled states to be transformed into their
respective orthogonal states. For non-maximally entangled states
(all of which are equally entangled), this covariance allows only
for {\it approximate} NOT operations. {\it None} of the operations,
we have considered in this work, satisfies this covariance, although
each of them is a unital trace-preserving CP map.

In this paper, we have noticed that so far as rotation of two-qubit
pure states, through the same amount by one-sided local quantum
operations, are concerned, the size of the maximal set of two-qubit
pure states will be {\it same} as that of the single-qubit pure
states, provided we restrict our attention to only equally entangled
two-qubit pure states.

So far, we have discussed only about LOCC (more specifically, local
operations only). In order to have an idea about the difference in
the sizes of the maximal sets ${\cal S}^{(l_0, l_1)}_{W_1 \otimes
W_2}$ and ${\cal S}^{(l_0, l_1)}_{W}$, respectively for local
rotator $W_1 \otimes W_2$ and non-local rotator $W$, let us consider
the swap operator $W_{swap} = (1/2)[I_2 \otimes I_2 + {\sigma}_x
\otimes {\sigma}_x + {\sigma}_y \otimes {\sigma}_y + {\sigma}_z
\otimes {\sigma}_z] \in U(4)$, so that $W \equiv e^{i\pi/4}W_{swap}
\in SU(4)$ is a non-local special unitary operator. Here we shall
use the fact that $W_{swap} =
2(|{\phi}^+{\rangle}_{AB}\!{\langle}{\phi}^+|)^{T_B}$, where
$|{\phi}^+\rangle_{AB} = (1/{\sqrt{2}})(|00\rangle_{AB} +
|11\rangle_{AB})$ and $T_B$ denotes partial transposition with
respect to the qubit $B$. Now ${\langle}{\psi}_0|(U^{\dagger}
\otimes V^{\dagger})W(U \otimes V)|{\psi}_0{\rangle} =~ {\rm Tr}
[(U^{\dagger} \otimes V^{\dagger})W(U \otimes
V)|{\psi}_0{\rangle}{\langle}{\psi}_0|] =~ {\rm Tr} [(U^{\dagger}
\otimes V^*)W^{T_B}(U \otimes
V^T)(|{\psi}_0{\rangle}{\langle}{\psi}_0|)^{T_B}] = 2e^{i\pi/4} {\rm
Tr} [(U^{\dagger} \otimes V^*)|{\phi}^+{\rangle}{\langle}{\phi}^+|(U
\otimes V^T)(|{\psi}_0{\rangle}{\langle}{\psi}_0|)^{T_B}] =
2e^{i\pi/4}{\langle}{\phi}^+|(U \otimes
V^T)\{(|{\psi}_0{\rangle}_{(l_0,
l_1)}\!{\langle}{\psi}_0|)^{T_B}\}(U^{\dagger} \otimes
V^*)|{\phi}^+{\rangle} = 2e^{i\pi/4}[l_0^2|{\langle}{\phi}^+|(U
\otimes V^T)|00{\rangle}|^2 + l_1^2|{\langle}{\phi}^+|(U \otimes
V^T)|11{\rangle}|^2 + l_0l_1|{\langle}{\phi}^+|(U \otimes
V^T)|{\psi}^+{\rangle}|^2 - l_0l_1|{\langle}{\phi}^+|(U \otimes
V^T)|{\psi}^-{\rangle}|^2]$, where $|{\psi}^{\pm}{\rangle} =
(1/{\sqrt{2}})(|01\rangle \pm |10\rangle)$. So, we are looking for
all possible $U, V \in SU(2)$ for which
$2e^{i\pi/4}[l_0^2|{\langle}{\phi}^+|(U \otimes V^T)|00{\rangle}|^2
+ l_1^2|{\langle}{\phi}^+|(U \otimes V^T)|11{\rangle}|^2 +
l_0l_1|{\langle}{\phi}^+|(U \otimes V^T)|{\psi}^+{\rangle}|^2 -
l_0l_1|{\langle}{\phi}^+|(U \otimes V^T)|{\psi}^-{\rangle}|^2] =
2e^{i\pi/4}[l_0^2|{\langle}{\phi}^+|00{\rangle}|^2 +
l_1^2|{\langle}{\phi}^+|11{\rangle}|^2 +
l_0l_1|{\langle}{\phi}^+|{\psi}^+{\rangle}|^2 -
l_0l_1|{\langle}{\phi}^+|{\psi}^-{\rangle}|^2]$, {\it i.e.},
$l_0^2|{\langle}{\phi}^+|(U \otimes V^T)|00{\rangle}|^2 +
l_1^2|{\langle}{\phi}^+|(U \otimes V^T)|11{\rangle}|^2 +
l_0l_1|{\langle}{\phi}^+|(U \otimes V^T)|{\psi}^+{\rangle}|^2 -
l_0l_1|{\langle}{\phi}^+|(U \otimes V^T)|{\psi}^-{\rangle}|^2 =
1/2$. Now the maximum value, the quantity
$l_0^2|{\langle}{\phi}^+|(U \otimes V^T)|00{\rangle}|^2 +
l_1^2|{\langle}{\phi}^+|(U \otimes V^T)|11{\rangle}|^2 +
l_0l_1|{\langle}{\phi}^+|(U \otimes V^T)|{\psi}^+{\rangle}|^2 -
l_0l_1|{\langle}{\phi}^+|(U \otimes V^T)|{\psi}^-{\rangle}|^2$ can
attain, is $1/2$, which occurs if $|{\langle}{\phi}^+|(U \otimes
V^T)|00{\rangle}|^2 = |{\langle}{\phi}^+|(U \otimes
V^T)|11{\rangle}|^2 = 1/2$ and $|{\langle}{\phi}^+|(U \otimes
V^T)|{\psi}^+{\rangle}|^2 = |{\langle}{\phi}^+|(U \otimes
V^T)|{\psi}^+{\rangle}|^2 = 0$. In this case $V$ (conversely, $U$)
can be chosen {\it arbitrarily} from $SU(2)$ but $U$ has to be
chosen from $SU(2)$ is such a way that $r^U_x = r^V_x$ and $r^U_y =
r^V_y$. Note that this is true {\it irrespective} of whether
$|{\psi}_0\rangle$ is maximally entangled or not. Thus we see that
the swap operator can not rotate, with equal amount, all the
maximally entangled states. Nevertheless, in the case of a
non-maximally entangled state $|{\psi}_0\rangle$, the maximal set
${\cal S}^{(l_0, l_1)}_{e^{i\pi/4}W_{swap}}$ is {\it much} larger
than the corresponding maximal set ${\cal S}^{(l_0, l_1)}_{W_1
\otimes I_2}$ (or ${\cal S}^{(l_0, l_1)}_{I_2 \otimes W_2}$) for
one-sided local unitary operators

As in the case of $2 \otimes 2$ systems, all maximally entangled
states of a $d \otimes d$ system can be rotated, through arbitrary
but same amount, by one-sided local unitary operators. The question
of rotating non-maximally entangled states of $d \otimes d$ is more
subtle as we don't have a Bloch sphere representation for $d$
dimensional quantum systems when $d > 2$. In fact, in the standard
product basis $\{|ij\rangle : i, j = 0, 1, \ldots, d - 1\}$ of
${C\!\!\!\!I}^d \otimes {C\!\!\!\!I}^d$, every normalized pure state
$|\psi\rangle = \sum_{i, j = 0}^{d - 1} c_{ij}|ij\rangle$ can be
represented by the $d \times d$ matrix $C \equiv (c_{ij})_{i, j =
0}^{d - 1}$ (where ${\rm Tr} [CC^{\dagger}] = 1$), or, equivalently,
by the symbol $|C{\rangle}\!{\rangle}$. Given the non-negative
diagonal matrix $D =~ {\rm diag} (D_0, D_1, \ldots, D_{d - 1})$
(with $0 \le D_{d - 1} \le D_{d - 2} \le \ldots D_0 \le 1$ and
$\sum_{i = 0}^{d - 1} D_i^2 = 1$), we would like to find out the
maximal set of all the equally entangled states $|\psi\rangle \equiv
(U \otimes V)|D{\rangle}\!{\rangle} = |UDV^T{\rangle}\!{\rangle} \in
{C\!\!\!\!I}^d \otimes {C\!\!\!\!I}^d$ (including the state
$|D{\rangle}\!{\rangle}$) for which ${\langle}\!{\langle}UDV^T|(W_1
\otimes W_2)|UDV^T{\rangle}\!{\rangle} = {\langle}\!{\langle}D|(W_1
\otimes W_2)|D{\rangle}\!{\rangle}$, {\it i.e.}, ${\rm Tr}
[W_1(UDV^T)W_2^T(UDV^T)^{\dagger}] =~ {\rm Tr} [W_1DW_2^TD]$, where
$W_1, W_2$ are fixed elements of $SU(d)$ while $U, V$ can be some or
all the elements of $SU(d)$. When $W_2 = I_d$, this condition
becomes ${\rm Tr} [W_1UD^2U^{\dagger}] =~ {\rm Tr} [W_1D^2]$, which
will be satisfied for all those $U \in SU(d)$, each of which
commutes with $W_1$ (but these are not the only solutions). When
$|D{\rangle}\!{\rangle}$ is a maximally entangled state, $D =
(1/{\sqrt{d}})I_d$, and so the last condition will be automatically
satisfied for this $D$ irrespective of the choice of $U$.

Every great circle ${\cal G}$ of the Bloch sphere (the state space
of any two-level quantum mechanical system) can be identified with
an element $W$ of $SU(2)$ in the sense that
${\langle}\psi|W|\psi{\rangle}$ is same only for the elements
$|\psi\rangle$ of ${\cal G}$. For example, the NOT operation
$i{\sigma}_y$ brings every element of the polar great circle ${\cal
G} = \{{\rm cos} ({\theta}/2)|0\rangle + {\rm sin}
({\theta}/2)|1\rangle | \theta \in [0, \pi]\} \bigcup \{{\rm cos}
({\theta}/2)|0\rangle -{\rm sin} ({\theta}/2)|1\rangle | \theta \in
[0, \pi]\}$ to its orthogonal state but it does not do the same job
for states lying outside of ${\cal G}$. The issue of finding out the
geometric structure of state space for higher dimensional quantum
systems is not yet resolved. So what would be the shape of the
boundary\footnote{The so called `great circle' of the corresponding
geometric structure of the state space.} of the intersection of this
geometric structure and any hyperplane passing through the centre of
this geometric body, is yet to be figured out, in general. Our
present work may through some light on the structure of this
boundary.



\end{document}